%% file: main.tex
\documentclass[11pt,letterpaper]{article} 
\usepackage{setspace,wrapfig}

\usepackage{mathptmx} 

\usepackage{mdframed}
\usepackage{bibspacing}
\input{structure.tex} 

\title{X-ray Studies of Exoplanets: \\
A 2020 Decadal Survey White Paper}

\author{\authorstyle{
\noindent Scott J. Wolk\textsuperscript{1}*, 
Jeremy J. Drake\textsuperscript{1}, 
Graziella Branduardi-Raymont\textsuperscript{2},
Katja Poppenhaeger\textsuperscript{3,4},
Vladimir Airapetian\textsuperscript{5},
Kevin France \textsuperscript{6},
Salvatore Sciortino\textsuperscript{7},
Ignazio Pillitteri\textsuperscript{7},
Rachel A. Osten\textsuperscript{8,9},
Carey M. Lisse\textsuperscript{9},
Vinay Kashyap\textsuperscript{1}, 
Brad Wargelin\textsuperscript{1},
Brian Wood\textsuperscript{10},
Willaim Dunn\textsuperscript{1},
David Principe\textsuperscript{11},
Moritz G\"unther\textsuperscript{11},
Damian J. Christian\textsuperscript{12},
Julián David Alvarado-Gómez\textsuperscript{1},
Chuanfei Dong\textsuperscript{13},
Lidia Oskinova\textsuperscript{3},
Margarita Karovska\textsuperscript{1},
Sofia P. Moschou\textsuperscript{1},
Peter K. Williams\textsuperscript{1},
Randall Smith\textsuperscript{1},
Bradford Snios\textsuperscript{1},
Elena Gallo\textsuperscript{14},
William Danchi\textsuperscript{15},
John P. Pye\textsuperscript{16},
Joel Kastner\textsuperscript{17},
Jose Dias Do Nascimento\textsuperscript{1,18},
Jae-Sub Hong\textsuperscript{19}
} 
\newline\newline 
\textsuperscript{1}\institution{Smithsonian Astrophysical Observatory}, 
\textsuperscript{2}\institution{University College London},
\textsuperscript{3}\institution{Queen's University Belfast, United Kingdom},
\textsuperscript{4}\institution{University of Potsdam, Germany},
\textsuperscript{5}\institution{NASA/GSFC \& American University},
\textsuperscript{6}\institution{JILA, University of Colorado},
\textsuperscript{7}\institution{INAF--Osservatorio Astronomico di Palermo},
\textsuperscript{8}\institution{Space Telescope Science Institute},
\textsuperscript{9}\institution{Johns Hopkins University},
\textsuperscript{10}\institution{Naval Research Laboratory},
\textsuperscript{11}\institution{Massachusetts Institute of Technology},
\textsuperscript{12}\institution{California State University Northridge},
\textsuperscript{13}\institution{Department of Astrophysical Sciences, Princeton University},
\textsuperscript{14}\institution{University of Michigan}, 
\textsuperscript{15}\institution{NASA Goddard Space Flight Center},
\textsuperscript{16}\institution{University of Leicester, United Kingdom},
\textsuperscript{17}\institution{Rochester Institute of Technology},
\textsuperscript{18}\institution{Univ. Federal do Rio G. do Norte, UFRN, Brazil},
\textsuperscript{19}\institution{Harvard University}
\textsuperscript{}\institution{*Corresponding author} 
}

\date{A White Paper submitted to the {\em Astro2020 decadal survey committee}, National Academies of Sciences, Engineering and Medicine\\ March 5, 2019}

\begin{document}

\maketitle 

\thispagestyle{firstpage} 


\newpage

\lettrineabstract{
Over the last two decades, the discovery of exoplanets has fundamentally changed our perception of the universe and humanity's place within it. Recent work indicates that a solar system's X-ray and high energy particle environment is of fundamental importance to the formation and development of the atmospheres of close-in planets such as hot Jupiters, and Earth-like planets around M stars. X-ray imaging and spectroscopy provide powerful and unique windows into the high energy flux that an exoplanet experiences, and X-ray photons also serve as proxies for potentially transfigurative coronal mass ejections. Finally, if the host star is a bright enough X-ray source, transit measurements akin to those in the optical and infrared are possible and allow for direct characterization of the upper atmospheres of exoplanets.  In this brief white paper, we discuss contributions to the study of exoplanets and their environs which can be made by X-ray data of increasingly high quality that are achievable in the next 10--15 years.}


\setcounter{page}{1}

\section{High Energy Photons in the Context of Exoplanets} 
The field of exoplanetary science is in transition from the discovery phase which marked its first two and a half decades to the detailed characterization of exoplanets hosted by stars across the Hertzprung-Russell (HR) diagram.  While some atomic species such as sodium have been detected in particular exoplanets, measurements of oxygen, ozone, and carbon dioxide, currently in their infancy, will transform the science in the coming decades.  Part of that characterization is 
necessary for the stellar ecosystem where the planet lives.  A star does not end at its photosphere; although more tenuous, its substance extends to surrounding high temperature coronal plasma, high energy flares, closed and open magnetic fields, stellar winds, and coronal mass ejections (CMEs).  In a very real sense, planets exist in the upper atmospheres of stars.
 
 Planetary characteristics, including habitability, are determined by much more than the mass of the exoplanet and the bolometric flux of the host star. Recent studies of atmospheric escape due to thermal and non-thermal processes concluded that close-in planets would lose mass proportional to the incident stellar X-ray and ultraviolet (XUV) flux (e.g.\ Lammer et al.\ 2003,  Erkaev et al.\ 2007, Sanz-Forcada et al.\ 2010).   The effective diameter of planets is larger at XUV wavelengths compared to optical wavelengths due to the absorption of low density gas at XUV energies. Further,  Roche lobe overflow plays a critical role in mass loss and the related Hill (tidal) radius is proportional to the -3 power of the orbital distance.  Analogously, magnetic interactions are proportional to the -3 power within the Alfv\'en radius, varying as the inverse square relation further out. The result is that close-in systems experience stronger influence by high energy particles and photons than na\"ively expected from scalings based simply on the stellar optical luminosity.
 
\begin{figure}[htb]
\centering 
\includegraphics[width=0.9\textwidth]{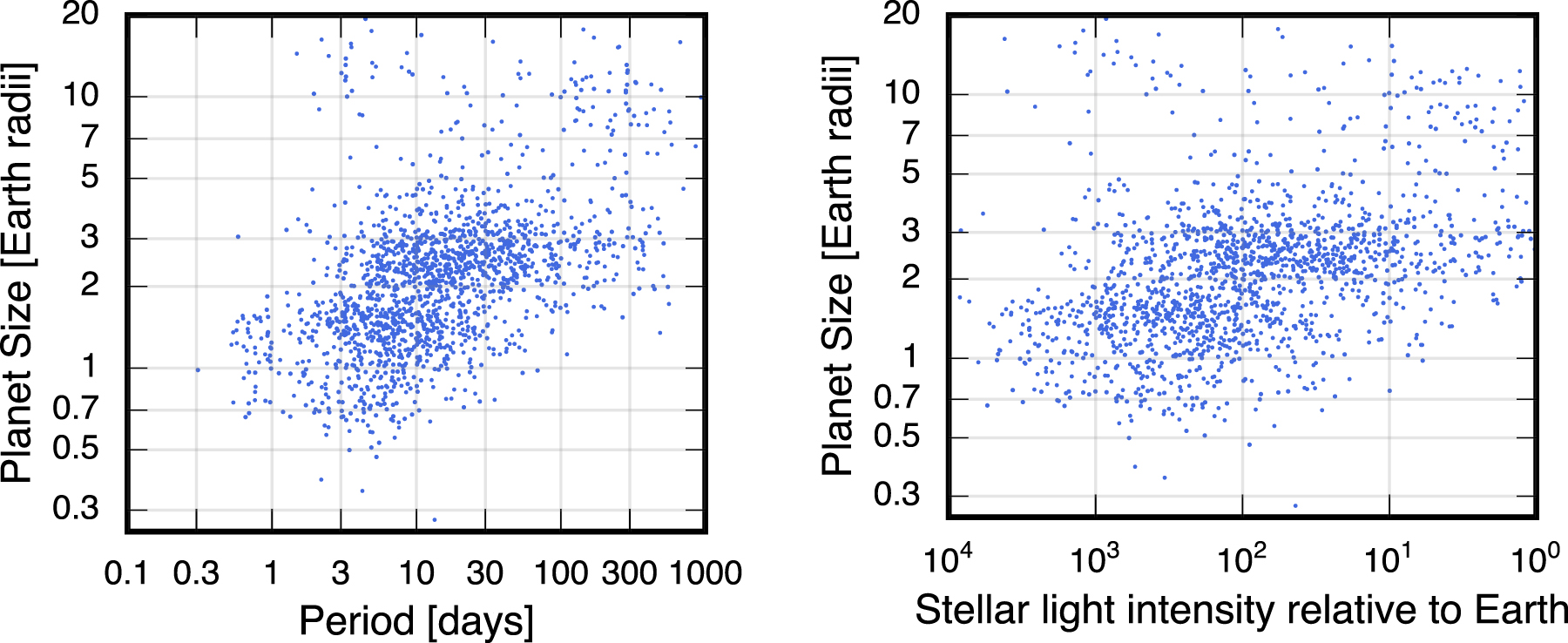}
\caption{
Left: distribution of planet radii and orbital periods. Right: same as left but with insolation flux relative to Earth on the horizontal axis. The dearth of exoplanets with periods less 3 days and radii between 2 and 20 Earth masses (the ``exoplanet evaporation desert'') is attributed to atmospheric losses due to high energy radiation (Fulton \& Petigura 2018).}
\label{f:OD2018}
\vspace{-0.0in}
\end{figure}
 
 High-energy photons likely drive the ``exoplanet evaporation desert'' (Figure \ref{f:OD2018}). In one recent example, {\it XMM-Newton} observations indicate WASP-19b could have already lost roughly half its current total mass in its $\sim$2.2 Gyr life due to incident X-ray flux (Lalitha et al.\ 2018). Statistically this effect has been shown by Kepler results (Fulton \& Petigura 2018) and described theoretically (Lopez \& Fortney 2013, Owen \& Wu 2017, Owen \& Lai 2018).


Much of the current and near term search for habitable zone planets has focused on M stars with relatively close-in planets.  This includes tantalizing cases such as TRAPPIST-1 (Gillon et al.\ 2017). Close-in planets lie in unique environments and the impact of the X-rays from the host star changes the evolution of the system. The effects can work several ways: \\
(1) the intense high energy flux heats the upper atmosphere of an exoplanet via photoionization, \\
(2) the angular momentum and magnetic field of the planet can induce more activity on the star, thereby leading to non-linear enhancements in the rates of heating, ionization and various atmospheric chemical processes,\\
(3) chemical reactions are also driven by particles absorbed in the upper atmosphere, and heating and dissociative reactions can lead  to erosion and evaporation of the  planetary atmospheres, particularly for close-in planets (Lammer et al. 2003, Yelle 2004, Tian et al. 2005, Murray-Clay et al. 2009, Airapetian et al. 2016, Dong et al. 2018).\\

In fact, atmospheric escape has already been observed in HD 189733b, HD 209458b, WASP-12b and GJ436b (Vidal-Madjar et al.2003, Linsky et al. 2010, Lecavelier Des Etangs et al. 2010, Fossati et al. 2010, Kulow et al. 2014, Ehrenreich et al. 2015). The kinetic impact of stellar coronal mass ejections (CMEs) and winds can be even more devastating, and penetrating particles can directly damage organics at the cellular level. For a full understanding of exoplanet atmospheres, their evolution, and  planetary habitability, it is therefore crucial that we characterize the host stars' high energy emission.

In the coming decades, we may be able to utilize X-rays to characterize exoplanets as we do planets in our own solar system.  Planetary auroral X-ray and radio emission can be used as a probe for difficult to detect planetary magnetic fields, which are thought to be vital for protecting a planet's atmosphere from high energy particle stripping.  Meanwhile, charge exchange between stellar winds and planetary atmospheres can be used as a probe of atmospheres and the interplanetary medium (Wargelin \& Drake 2001, Cravens et al.\ 2001, Bhardwaj et al.\ 2007).  Advances in our understanding will come from a combination of statistical studies and focused investigations of carefully selected interesting systems.  These will be chosen from the thousands of extra-solar planets discovered and explored by the {\it CoRoT}, {\it Kepler}, {\it TESS}, {\it CHEOPS},  {\it ARIEL}, and {\it PLATO} space missions, while {\it Gaia} will help to identify the young, highly eccentric systems  best suited to test models of UV and X-ray irradiation effects as a function of star-planet separation and orbital phase.

\section{Characterization of the High Energy Stellar Flux of Exoplanet Hosts} 

Stellar coronal activity and flaring are ubiquitous in solar-type and lower-mass stars, and X-ray observations can be used to determine the temperature, metallicity, and electron density of coronal plasma.  Previous studies indicate significant secular decrease of stellar XUV flux with age. Enhanced X-ray emission and higher temperatures during early epochs means that X-rays at early times will dissociate and ionize molecules in planetary thermospheres and exospheres more readily (G\"udel 2007; Penz, Micela \& Lammer 2008). It is not clear if this has a deleterious result or is a necessary mechanism of clearing extremely dense proto-atmospheres. 

Additional energy available due to star--planet interaction (SPI) is expected to roughly scale with the strength of the individual magnetic fields ($B_*$ and $B_P$), the relative velocity of the star and planet ($v_{rel}$), and their separation ($a$) according to $B_* B_P v_{rel} a^{-n}$ (Cuntz et al. 2000, Lanza 2009, 2012, Iro \& Demming 2010). While the exponential $n$ is $\sim$2 in the open-field region of the stellar wind,  $n$ is $\sim$ 3 when close to the star where the field is dipole-like. It is possible that such an effect could be detected by statistical studies of large well chosen samples.  Indeed this was attempted using {\it ROSAT} All Sky Survey (RASS) data; a weak dependence between X-ray flux and semi-major axis was noted (Kashyap et al. 2008).  More recent work finds some evidence of measurable X-ray flux excesses in close-in systems, but this is dominated by a few extreme systems where SPI seems to play a significant role (Miller et al.\ 2015). Other phenomena related to SPI include stellar spin-up effects (Poppenhaeger \& Wolk 2014), dynamo quenching (or obscuration) (Pillitteri et al. 2014, Fossati et al. 2018),
and coronal abundance alterations (Wood et al. 2018).

Current studies are severely constrained by instrumental limitations. {\it Chandra} has limited effective area, and {\it XMM-Newton} has modest spatial resolution, reducing its ability to study young stars in clusters. Both facilities use Si-based X-ray detectors with imaging spectral resolution of $R=\frac {E}{\Delta E}  \approx 50$. In addition to surveys, extended monitoring of specific interesting sources is needed to understand the range of conditions planets may be subjected to; impulsive events are particularly easy to miss in surveys.

While major observational advances will require a combination of more sensitive detectors and larger collecting area,  substantial progress in a few areas can be made with focused missions using current technology.  Surveys of older, more isolated stars could benefit substantially from modest spatial resolution X-ray instruments. The RASS revealed about 50,000 normal stars with fluxes above $2 \times 10^{-13}$erg cm$^{-2}$s$^{-1}$ in the soft X-ray band.  
The energy resolution of modern detectors is $R \approx 100$, making them much less sensitive to panchromatic background noise.  Short observations of a few hours could measure a luminosity and temperature for an exoplanet host star in the RASS. 

For the brightest X-ray sources, the effective area requirement is low and it is possible to follow-up with specialized low cost small X-ray satellites which can monitor exoplanet hosts for flares (and transits).  Specifically, Pillitteri et al. (2011, 2014) have suggested that some of the activity observed on HD 189733A is tied to the planetary orbital period and not the stellar rotation period. Similarly, Maggio et al.\ (2015) detected flaring in HD 17156 at the periastron passage of its highly eccentric exoplanet.  Currently, reproducing such observations requires devoting scarce "Great Observatory" scale resources to monitoring for a stochastic signal, which then needs to be compared to an extended baseline to understand its significance. Given a focused long term monitoring program and using a devoted low cost mission it will be possible to confirm the relationship between the synoptic flares and differentiate them from intrinsic stellar events, primarily on the basis of timing.

Further advances in distinguishing activity induced by star--planet interactions from  intrinsic stellar activity, will require more effective area coupled with higher spectral resolution (R$\sim$ 500--2000).  This can be done using either micro-calorimeters or dispersive optics.  Such an experiment would be able to collect more photons in less time and thus spectrally resolve ephemeral events. With this moderate spectral resolution, coupled with collecting area on the order of 1000-10,000 cm$^2$, we will be able to spectrally resolve flares on the time scale of minutes.  Work on young stars and planets, which are predicted to be the most affected by such events, will require excellent spatial resolution as well; to perform background limited studies of individual sources in the densest regions of even the closest massive young star forming cluster (Orion), spatial resolution on the order of 1$^{\prime\prime}$ would be required.

Strong X-ray flares on the Sun are usually accompanied by the ejection of cooler material (roughly 10,000 K) that had previously been confined by magnetic fields near the solar surface.  These coronal mass ejections (CMEs) may also contain high energy protons accelerated in the flare and  CME shock front. CMEs differ from the quasi-steady solar wind in two respects: they are orders of magnitude denser, and are spatially confined.  CMEs have been theorized to induce exotic chemical reactions in planetary atmospheres which may be required for formation of complex molecular compounds, such as amino acids (e.g., Airapetian et al.\ 2016, Lingam et al.\ 2018). Conversely, CMEs are a threat to exoplanet habitability because their impacts can, over time, strip off a significant fraction of a planet's atmosphere, especially its ozone. With similar collecting area but even higher resolution spectroscopy (R$\sim$ 5000), X-ray astronomers could routinely and definitively observe the tell-tale Doppler shifts of CMEs or their coronal compression waves and measure their physical properties, including velocity, temperature, mass, and energy.

\section{Transits}

Remarkable progress has been made in studying the atmospheres of hot Jupiters by observing transits in the optical and IR. Additional and unique insights can be obtained via X-ray transits, although the much lower photon fluxes and greater intrinsic variability mean that repeated observations will be necessary.  In the UV, higher energy measurements allow access to layers of the atmosphere lying above the optical photosphere.  In the case of UV this is limited to about 10\% above the optical cloud tops.  X-rays are absorbed at far lower densities and hence can resolve details up to twice the altitude of the optical clouds  (Vidal-Madjar et al.\ 2003, Linsky et al.\ 2010). Lecavelier des Etangs et al.\ (2012) ascribe their detection of Ly${\alpha}$ transits to X-ray flares, suggesting a role for X-ray observations in coordinated multiwavlength observations with JWST and ALMA. More recently, Lavie et al.\ (2017) have detected an enlarged asymmetric exosphere around the hot Neptune GJ 436b.

\begin{figure}[htb]
\centering 
\includegraphics[width=0.9\textwidth]{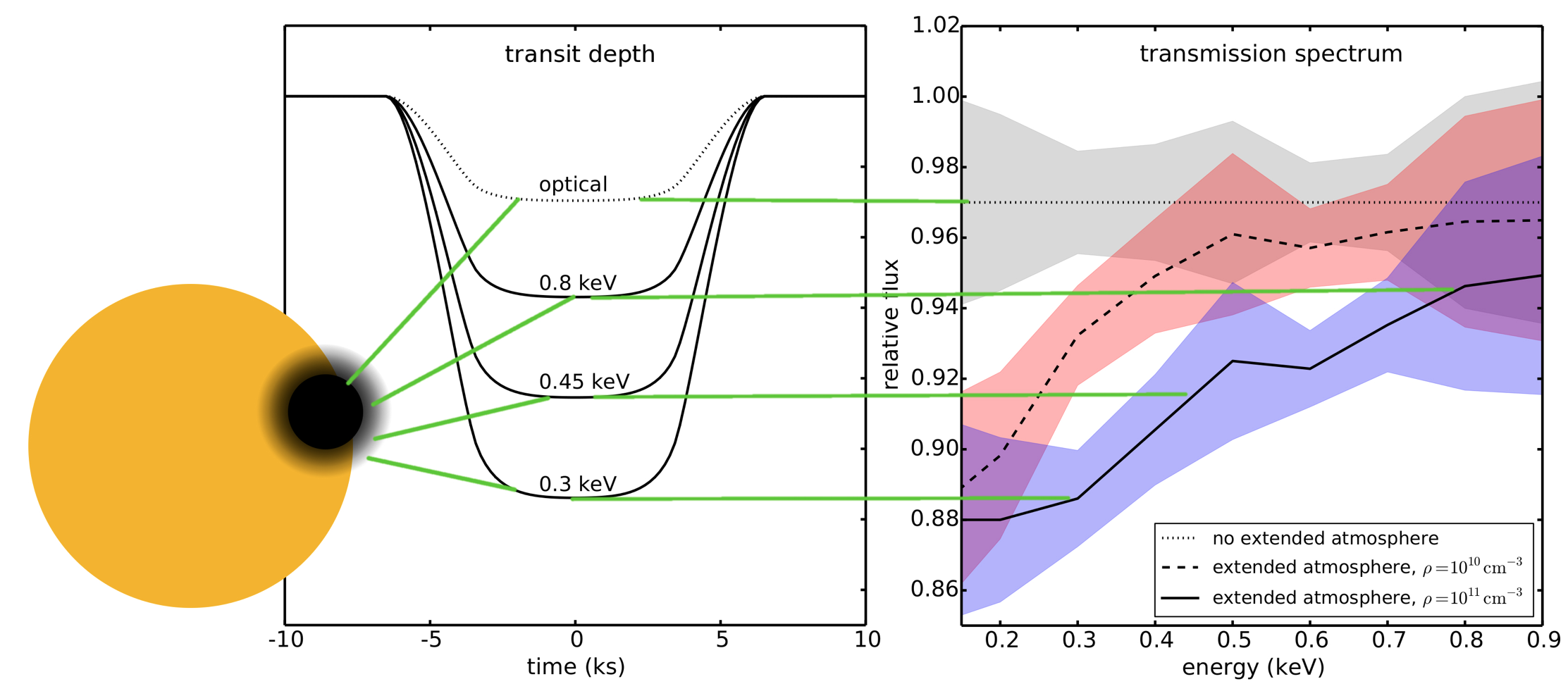}
\caption{Left: Representation of the relative depth of light curves observed at various energies due to different absorption cross-sections.  Right: ~~ Simulation of detection of the 0.5 keV oxygen absorption edge betraying enhanced O abundance for multiple transits of a super-earth planet around an M dwarf  (by K. Poppenhaeger).}
\label{f:3color_transit}
\vspace{-0.0in}
\end{figure}
 
Using {\it Chandra}, Poppenhaeger et al.~(2013) monitored multiple transits of HD~189733b. After excluding flares and co-adding the remaining lightcurves, they were able to measure the transit width and its depth (7\%), much deeper than the $\approx~2\%$ optical eclipse depth. This implied a soft  X-ray ($<$1 keV) planet radius of about 1.75 times the optical size.  They identified the source of the additional obscuration as due to C,N,O in a spherically symmetric exosphere with a density of about $10^{11}$ cm$^{-3}$ and a temperature of about 20,000K.  Future X-ray observations could potentially measure asymmetries in transit  lightcurves, pointing to stellar wind induced atmospheric depletion, comet-like sublimation tails,  or bow shocks around planets.  Using hot-Jupiter observations and modeling as a guide, one could then constrain the loss of 
atmospheres in exoEarths around M-stars (Figure~\ref{f:3color_transit}).
 
The more photons that arrive during the transit the greater the number of samples along the light curve and the higher precision that is possible in each individual sample or temporal bin. Statistically, we expect a few dozen {\it TESS}-revealed exoplanet hosts to be detected in the RASS, which has a limiting flux of $\approx$ 2 $\times 10^{-13}$ erg cm$^{-2}$ sec$^{-1}$.  For these sources, about 10,000 cm$^2$ is required to achieve a rate of 1 cnt sec$^{-1}$. For a single time bin to have a precision of 10\% would require a minimum of 100 counts, or 10$^4$ counts for
a precision of 1\%.
 
 
 The typical transit duration for a close-in system is 2--3 hours. In principle, a single transit could reveal a small fluctuation, but in practice, repeated observations are needed because of fluctuations of active regions on the stellar surface (c.f.\ Llama \& Shkolnik 2015).  In addition, while the detection of a small dip due to a transit is useful, the key science of atmospheric modeling comes from the ability to measure the shape of the light curve.  Multiple transit observations allow the observer to correct for the underlying variations and allow finer time bins. Given the 1000-10,000 cm$^2$ described above, 5-10 transit observations would allow for the creation of a high resolution light curve with error bars approaching a few percent. If dispersive gratings were used various ionic species could be sorted by altitude (Figure~2).  Alternatively, a dedicated small telescope with an effective area on the order 100 cm$^2$ could measure exospheric struture by co-adding 50-100 transits.  
 
 Conversely, given greater collecting area ($\approx 20,000~$cm$^2$) and better spatial resolution (to reduce noise, especially from cosmic rays and secondary astrophysical sources in the field), the coming decades will allow us to further investigate the vertical distribution of atmospheric species.  Extremely high resolution maps of individual planetary atmospheres are possible.  The transverse velocity of a close-in exoplanet is around 150 km/s.  Poppenhaeger \& Wolk (2014) found the X-ray diameter of HD 189733b at around 800 eV is about 1.75 times the optical diameter or nearly 250,000 km.  The diameter should be even larger than this at lower energies near the carbon edge ($\approx 300$ eV) where the opacity per particle is much higher. By combining photons from multiple transits and concentrating on photons within narrow energy bands corresponding to (e.g.) oxygen or carbon it will be possible to map structures on scales perhaps 5--10\% the planetary diameter.

\section{Summary}
The coming decades will open a new opportunity of exploring the habitability of other worlds.  X-ray observations of exoplanet hosts will be critical to this understanding.  Devoted X-ray instruments of modest size and resolution will be able to contribute.  They will be especially useful in monitoring and transit studies of bright exoplanets and their hosts.  However, the most vital scientific questions will require large effective area and spatial and spectral resolutions sufficient to distinguish stars, elemental species, and dynamic structures at a variety of scales. 
\bigskip 

\eject
\small
\bibliographystyle{aasjournal}


{\Large\usefont{OT1}{phv}{b}{n}\color{Black} \noindent References}

\input{Exo_refs-archive}



\end{document}

%% file: structure.tex
\usepackage[english]{babel} 

\usepackage{microtype} 

\usepackage{amsmath,amsfonts,amsthm} 

\usepackage[svgnames]{xcolor} 

\usepackage[hang, small, labelfont=bf, up, textfont=it]{caption} 

\usepackage{booktabs} 

\usepackage{lastpage} 

\usepackage{graphicx} 

\usepackage{enumitem} 
\setlist{noitemsep} 

\usepackage{sectsty} 
\allsectionsfont{\usefont{OT1}{phv}{b}{n}} 

\usepackage{geometry} 

\usepackage[T1]{fontenc} 
\usepackage[utf8]{inputenc} 

\usepackage{XCharter} 

\usepackage{fancyhdr} 
\fancypagestyle{firstpage}{ 
	\fancyhf{}
}

\newcommand{\authorstyle}[1]{{\large\usefont{OT1}{phv}{b}{n}\color{MidnightBlue}#1}} 

\newcommand{\institution}[1]{{\small\usefont{OT1}{phv}{m}{sl}\color{Black}#1}} 

\usepackage{titling} 

\newcommand{\HorRule}{\color{DarkGoldenrod}\rule{\linewidth}{1pt}} 

\pretitle{
	\vspace{-50pt} 
	\HorRule\vspace{-30pt} 
	\fontsize{30}{36}\usefont{OT1}{phv}{b}{n}\selectfont 
	\color{MidnightBlue} 
\begin{flushleft}
}

\posttitle{\par\end{flushleft}\vskip 10pt} 

\preauthor{\begin{flushleft}} 

\postauthor{ 
\end{flushleft}
	\vspace{10pt} 
	\par\HorRule 
	\vspace{10pt} 
}

\usepackage{lettrine} 
\usepackage{fix-cm}	

\newcommand{\initial}[1]{ 
	\lettrine[lines=3,findent=4pt,nindent=0pt]{
		\color{DarkGoldenrod}
		{#1}
	}{}%
}

\usepackage{xstring} 

\newcommand{\lettrineabstract}[1]{
	\StrLeft{#1}{1}[\firstletter] 
	\initial{\firstletter}\textbf{\StrGobbleLeft{#1}{1}} 
}

\usepackage{natbib}

\newcommand\aj{{AJ}}
\newcommand\apj{{ApJ}}
\newcommand\apjl{{ApJL}}     
\newcommand\aap{{A\&A}}
\newcommand\mnras{{MNRAS}}
\newcommand\nat{{Nature}}
\newcommand\jgr{{JGR}}

%% file: Exo_refs-archive.tex
{\small

\noindent 
Airapetian, V.~S., Glocer, A., Gronoff, G., H{\'e}brard, E., \& Danchi, W.\ 2016, Nature Geoscience, 9, 452\\

\noindent 
 Bhardwaj, A., et al.\ 2007, Planetary and Space Science, 55, 1135\\

\noindent 
Cravens, T.~E., Robertson, I.~P., \& Snowden, S.~L.\ 2001, \jgr, 106, 24883 \\

\noindent 
Cuntz, M., Saar, S.~H., \& Musielak, Z.~E.\ 2000, \apjl, 533, L151\\

\noindent 
Dong, C.~F., Jin, M., Lingam, M., Airapetian, V. S., Ma, Y. J., van der Holst, B.\ 2018, Proc. Natl. Acad. Sci., 115, 260\\

\noindent 
Ehrenreich, D., et al.\ 2015, \nat, 522, 459 \\

\noindent 
 Erkaev, N.~V., Kulikov, Y.~N., Lammer, H., Selsis, F., Langmayr, D., Jaritz, G.~F., \& Biernat, H.~K.\ 2007, \aap, 472, 329 \\

\noindent
Fossati, L., et al.\ 2010, \apjl, 714, L222 \\

\noindent
Fulton, B.~J., \& Petigura, E.~A.\ 2018, \aj, 156, 264\\

\noindent
Gillon, M., et al.\ 2017, \nat, 542, 456 \\

\noindent
 G{\"u}del, M.\ 2007, Living Reviews in Solar Physics, 4, 3 \\

\noindent
Iro, N., \& Deming, L.~D.\ 2010, \apj, 712, 218 \\

\noindent
Kashyap, V.~L., Drake, J.~J., \& Saar, S.~H.\ 2008, \apj, 687, 1339\\

\noindent
Kulow, J.~R., France, K., Linsky, J., \& Loyd, R.~O.~P.\ 2014, \apj, 786, 132 \\

\noindent
Lalitha, S., Schmitt, J.~H.~M.~M., \& Dash, S.\ 2018, \mnras, 477, 808 \\

\noindent
Lammer, H., Selsis, F., Ribas, I., Guinan, E.~F., Bauer, S.~J., \& Weiss, W.~W.\ 2003, \apjl, 598, L121 \\

\noindent
Lanza, A.~F.\ 2009, \aap, 505, 339 \\

\noindent
 Lanza, A.~F.\ 2012, \aap, 544, A23 \\

\noindent
Lavie, B., et al.\ 2017,  \aap, 605, L7 \\

\noindent
 Lecavelier des Etangs, A., et al.\ 2012, \aap, 543, L4 \\

\noindent
Lingam, M., Dong, C. F., Fang, X. H., Jakosky, B. M., Loeb, A.\ 2018, \apj, 853, 10 \\

\noindent
 Linsky, J.~L., Yang, H., France, K., Froning, C.~S., Green, J.~C., Stocke, J.~T., \& Osterman, S.~N.\ 2010, \apj, 717, 1291 \\

\noindent
 Llama, J., \& Shkolnik, E.~L.\ 2015, \apj, 802, 41 \\
 
\noindent
Lopez, E.~D., \& Fortney, J.~J.\ 2013, \apj, 776, 2 \\

\noindent
Maggio, A., et al.\ 2015, \apjl, 811, L2 \\

\noindent
Miller, B.~P., Gallo, E., Wright, J.~T., \& Pearson, E.~G.\ 2015, \apj, 799, 163 \\

\noindent
Murray-Clay, R.~A., Chiang, E.~I., \& Murray, N.\ 2009, \apj, 693, 23 \\

\noindent
Owen, J.~E., \& Wu, Y.\ 2017, \apj, 847, 29 \\

\noindent
Owen, J.~E., \& Lai, D.\ 2018, \mnras, 479, 5012 \\

\noindent
Penz, T., Micela, G., \& Lammer, H.\ 2008, \aap, 477, 309 \\

\noindent
Pillitteri, I., G{\"u}nther, H.~M., Wolk, S.~J., Kashyap, V.~L., \& Cohen, O.\ 2011, \apjl, 741, L18 \\

\noindent
Pillitteri, I., Wolk, S.~J., Lopez-Santiago, J., G{\"u}nther, H.~M., Sciortino, S., Cohen, O., Kashyap, V., \& Drake, J.~J.\ 2014, \apj, 785, 145 \\

\noindent
Poppenhaeger, K., Schmitt, J.~H.~M.~M., \& Wolk, S.~J.\ 2013, \apj, 773, 62 \\

\noindent
Poppenhaeger, K., \& Wolk, S.~J.\ 2014, \aap, 565, L1 \\

\noindent
Sanz-Forcada, J., Ribas, I., Micela, G., Pollock, A.~M.~T., Garc{\'{\i}}a-{\'A}lvarez, D., Solano, E., \& Eiroa, C.\ 2010, \aap, 511, L8 \\

\noindent
 Tian, F., Toon, O.~B., Pavlov, A.~A., \& De Sterck, H.\ 2005, Science, 308, 1014 \\

\noindent
Vidal-Madjar, A., Lecavelier des Etangs, A., D{\'e}sert, J.-M., Ballester, G.~E., Ferlet, R., H{\'e}brard, G., \& Mayor, M.\ 2003, \nat, 422, 143\\

\noindent
Wood, B.~E., Laming, J.~M., Warren, H.~P., \& Poppenhaeger, K.\ 2018, \apj, 862, 66 \\

\noindent
Yelle, R.~V.\ 2004, Icarus, 170, 167 \\
}